# Influence of Rhenium Concentration on Charge Doping and Defect Formation in MoS$_2$

Kyle T. Munson,[1†] Riccardo Torsi,[1] Fatimah Habis,[2] Lysander Huberich,[3] Yu-Chuan Lin,[1] Yue Yuan,[4] Ke Wang,[5] Bruno Schuler,[3] Yuanxi Wang,[2] John B. Asbury,[1,4*] Joshua A. Robinson[1,4,5,6#]

1. Department of Materials Science and Engineering, The Pennsylvania State University, University Park, Pennsylvania 16802, United States
2. Department of Physics, University of North Texas, Denton, Texas 76203, United States
3. nanotech@surfaces Laboratory, Empa-Swiss Federal Laboratories for Materials Science and Technology, Dübendorf 8600, Switzerland
4. Department of Chemistry, The Pennsylvania State University, University Park, Pennsylvania 16802, United States
5. Materials Research Institute, The Pennsylvania State University, University Park, PA, 16802, USA
6. Department of Physics, The Pennsylvania State University, University Park, Pennsylvania 16802, United States

† kmm7044@psu.edu

* jasbury@psu.edu

# jar403@psu.edu

## Abstract

Substitutionally doped transition metal dichalcogenides (TMDs) are the next step towards realizing TMD-based field effect transistors, sensors, and quantum photonic devices. Here, we report on the influence of Re concentration on charge doping and defect formation in MoS$_2$ monolayers grown by metal-organic chemical vapor deposition. Re-MoS$_2$ films can exhibit reduced sulfur-site defects; however, as the Re concentration approaches $\gtrapprox$ 2 atom%, there is significant clustering of Re in the MoS$_2$. *Ab Initio* calculations indicate that the transition from isolated Re atoms to Re clusters increases the ionization energy of Re dopants, thereby reducing Re-doping efficacy. Using photoluminescence spectroscopy, we show that Re dopant clustering creates defect states that trap photogenerated excitons within the MoS$_2$ lattice. These results provide insight into how the local concentration of metal dopants affect carrier density, defect formation, and exciton recombination in TMDs, which can aid the development of future TMD-based devices with improved electronic and photonic properties.

**Introduction**

Monolayer transition metal dichalcogenide (TMD) semiconductors such as $MoS_2$ are appealing candidates for next-generation optoelectronic devices due to their direct bandgap,[1] large surface area to volume ratio,[2] stable excitons at room temperature,[3] and high carrier mobilities.[4, 5] The future use of TMDs in transistor, light emitting diode, and quantum photonic applications requires the ability to tune the electronic and photonic properties of TMDs using controlled doping methods.[6] Efforts to understand and control doping in TMDs focus on the influence of substrates,[7-9] interactions with adsorbed molecular species,[10-13] and the substitutional doping of foreign atoms at TMD metal or chalcogen sites.[14, 15] Of these approaches, substitutional metal doping offers the most viable means of incorporating stable dopants into the TMD lattice.[16] However, the impact of substitutional doping on carrier density and photonic properties in TMDs has been mixed and may be complicated by non-uniform dopant densities and the high ionization energies of some metal dopants.[16-18]

Substitutional doping is commonly achieved in techniques such as solid-source chemical vapor deposition by vaporizing powders or liquid precursors containing p- or n-type metal dopants during TMD growth.[16, 19-22] However, TMDs synthesized using these methods often exhibit non-uniform dopant concentrations and poor spatial uniformity.[22, 23] Additionally, the ionization energy of dopants in two-dimensional (2D) materials is higher than their bulk analogs due to quantum confinement effects and reduced dielectric screening at the monolayer level.[18, 24] As a result, carrier doping from metal dopant atoms is inefficient, prompting the use of > 1% dopant concentrations to tune the electronic properties of TMDs.[17, 25-27] Dopant clustering and dopant-dopant interactions are expected to be prevalent when the concentration of metal dopants reaches the levels identified to enable carrier doping of TMDs.[17, 28, 29] In classical semiconductors, increased interactions between dopant atoms at dopant concentrations > 0.01% can create electronically inactive dopant centers or impurity-related bands within the material's bandgap.[30-32] However, the impact of local dopant concentration on the structural and optoelectronic properties of substitutionally doped TMD monolayers remains an open area of research.

Uniform, electronic-grade TMDs with controlled dopant densities were recently synthesized from gas phase precursors using metal-organic chemical vapor deposition (MOCVD).[33, 34] In particular, MOCVD grown single-layer $MoS_2$ films doped with 0.05 to 1.0 atom% Re atoms[35] were shown to reduce the density of sulfur vacancy defects due to favorable dopant-defect interactions at the growth front of $MoS_2$ grains during growth.[35] Additionally, $Re_{Mo}$ in the negative, neutral, and positive charge state could be stabilized due to the closely-space donor state manifold.[36] The reduction of sulfur vacancy density and increased electron density following Re doping helped to improve electron transport in back gated field-effect transistors and reduce emission from defect states within $Re-MoS_2$.[35]

In this work, we demonstrate that local concentration variations of Re dopants can have a pronounced effect on charge doping and defect formation in $MoS_2$. Verified by X-ray photoelectron spectroscopy (XPS) and laser ablation inductively coupled plasma mass spectrometry (LA-ICP-MS), we demonstrate that MOCVD enables controllable introduction of low ($\lessapprox$ 1 atom%) to high ($\gtrapprox$ 1 atom%) Re concentrations in $MoS_2$ monolayers. Z-contrast scanning transmission electron microscopy (Z-STEM) measurements reveal that Re doping reduces the density of sulfur-site defects in $MoS_2$ over a range of dopant concentrations up to 6 atom%. However, valance band maximum (VBM), Raman, and photoluminescence (PL) measurements demonstrate that $Re-MoS_2$ films doped with high concentrations of Re atoms exhibit reduced n-type doping, increased strain, and broad sub-bandgap emission from Re-based defect states. Using a combination of scanning tunneling microscopy (STM) and *ab initio* calculations, we show that Re clustering at high dopant concentrations is responsible for the observed behavior.

**Results and Discussion**

$Re-MoS_2$ films (**Figure S1**) are synthesized onto c-plane sapphire and quasi-free standing epitaxial graphene (QFEG) substrates at 1000 °C using a multi-step MOCVD growth process previously described.[33-35] This process utilizes separate nucleation, ripening, and lateral growth stages to produce coalesced, monolayer $MoS_2$ films (**Figure S1**).[35, 37] Re dopants are incorporated into the $MoS_2$ lattice by flowing $Re_2(CO)_{10}$ during the growth process via a mass flow controller. The relationship between Re concentration and $Re_2(CO)_{10}$ flow rate is highlighted in **Figure S2**. The Re concentration (**Figure S2**) is obtained by XPS measurements of Re $4f_{7/2}$ and Re $4f_{5/2}$ peaks and LA-ICPMS (**Figure S2**). These results demonstrate that the MOCVD process can systematically tune the average Re concentration in $MoS_2$ from low to high doping percentages, where we define the boundary of low-to-high concentration as 1 atom%.

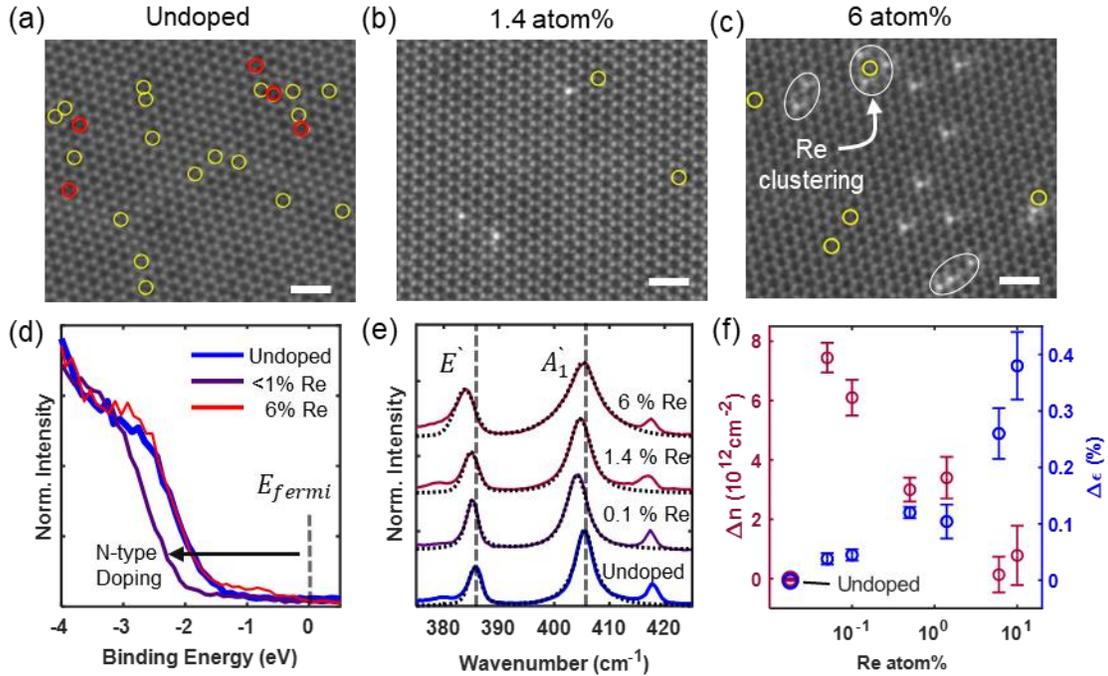

**Figure 1. Influence of Re concentration on structural disorder and n-type doping in Re-MoS$_2$:** Atomic resolution STEM images of (a) undoped, (b) 1.4, and (c) 6 atom% Re-MoS$_2$ films highlighting sulfur-site defects (yellow circles) and double sulfur-site defects (red circles). The scale bar is 1 nm. The images show reduced sulfur-site defects in Re-MoS$_2$ films. Re clustering is observed in 6 atom% films. (d) VBM edges of undoped, 0.1, and 6 atom% Re-MoS$_2$ films measured by XPS. (e) Raman spectra of undoped, 0.1, 1.4, and 6 atom % Re-MoS$_2$ films. The film's $A_1^`$ and $E^`$ modes are fit with pseudo-Voight functions to characterize their line widths and positions. The vibrational mode at ~417 cm$^{-1}$ is due to the sapphire substrate. (f) Raman-derived changes in strain ($\Delta\varepsilon$) and charge doping ($\Delta n$) as a function of Re concentration. The error bars indicate the standard deviation of ten Raman measurements collected across the samples.

Rhenium doping impacts the structural properties of MoS$_2$. This is evident when comparing Z-STEM images (**Figure 1a-c** and **Figure S3**) of undoped, 1.4, and 6 atom% Re-MoS$_2$ films. Analysis of the Z-STEM images reveals reduced sulfur vacancies (yellow circles) and double sulfur vacancies (red circles) in Re-MoS$_2$ films (**Figures 1a-c**). We acknowledge that the sulfur vacancies observed in our Z-STEM images may be filled with oxygen or carbohydrate species,[35, 38-41] therefore, we refer to these vacancies generally as sulfur-site defects. Sulfur-site defect reduction has been reported previously in MoS$_2$ films doped with dilute amounts of Re ($\leq$ 1 atom%). This defect reduction is attributed to Re atoms increasing the formation energy of sulfur vacancies at the growth front of MoS$_2$ grains.[35] The Z-STEM images shown in **Figure 1** highlight that these favorable dopant-defect interactions persist at higher doping concentrations up to 6 atom%. However, while the 1.4 atom% film exhibits isolated Re dopants, the 6 atom% Re-MoS$_2$ film exhibits non-uniform Re dopant clustering. The circled regions in the STEM image **(Figure**

**1c)** highlight this clustering. More examples of regular clustering patterns are discussed later in STM measurements. Furthermore, at 10 atom% doping levels, we observe Re dopant aggregation into ~3 nm phase-segregated domains whose crystal structure closely resembles that of ReS$_2$ (**Figure S4**).[42]

The Re dopant concentration modifies charge carrier doping within Re-MoS$_2$ monolayers. Analysis of the electron binding energy at the XPS measured valance band maximum (VBM) provides direct evidence of n-type doping due to Re incorporation.[21] The VBM edges of undoped, < 1 atom%, and 6 atom% Re-MoS$_2$ films are shown in **Figure 1d**. Evaluation of the VBM positions shows that only Re-MoS$_2$ films doped with low concentrations of Re atoms (<1 atom%) exhibit a VBM shift consistent with n-type doping.[21, 35] Conversely, samples doped with 6 atom% Re do not exhibit a shift in VBM position relative to undoped MoS$_2$, indicating a lack of n-type doping from Re atoms in highly doped films. Raman measurements correlate well with reduced n-type doping and increased strain in highly doped Re-MoS$_2$ films. The Raman spectra of undoped, 0.1, 1.4, and 6 atom% Re-MoS$_2$ films collected following 532 nm excitation are shown in **Figure 1e**. The spectra exhibit characteristic in-plane ($E^`$ ~ 385 cm$^{-1}$) and out-of-plane ($A_1^`$ ~ 405 cm$^{-1}$) vibrational modes for monolayer MoS$_2$.[7, 43] The vibrational feature at ~417 cm$^{-1}$ is due to the underlying sapphire substrate. The effect of strain and charge doping on the Raman modes of MoS$_2$ is well established in the literature.[7, 44-47] For MoS$_2$, the relationship between biaxial strain (ε), charge doping, and $E^`$ and $A_1^`$ peak positions is given by,[7]

$$\Delta\omega_E = -2\gamma_E \omega_0^E \varepsilon + k_n^E n \quad \text{(Eqn. 1)}$$

$$\Delta\omega_A = -2\gamma_A \omega_0^A \varepsilon + k_n^A n \quad \text{(Eqn. 2)}$$

where $\omega_0^E$ and $\omega_0^A$ are the frequencies of the MoS$_2$ $E^`$ and $A_1^`$ modes in the absence of strain and doping, $\gamma_E$ and $\gamma_A$ are Grüneisen parameters equal to 0.86 and 0.15 for the $E^`$ and $A_1^`$ modes,[45] $n$ is electron concentration in units of $10^{13}$ cm$^{-2}$, and $k_n^E$ and $k_n^A$ describe how charge doping shifts the $E^`$ and $A_1^`$ modes according to,[7]

$$k_n^E = -0.33 \frac{cm^{-1}}{10^{13}/cm^2} \quad \text{and} \quad k_n^A = -2.22 \frac{cm^{-1}}{10^{13}/cm^2}.$$

From Eqns. 1–2, the effect of Re concentration on strain and charge doping can be estimated by comparing the Raman peak positions of Re-MoS$_2$ films to undoped MoS$_2$ (see Supporting Information). Changes in lattice strain and charge doping as a function of Re atom% obtained from analysis of the Raman spectra (**Figure 1e**) are displayed in **Figure 1f**. The data show that strain ($\Delta\varepsilon$) within the MoS$_2$ lattice increases by ~ 0.3 % as the concentration of Re dopants increases from 0 (undoped) to 6 atom%. Additionally, the charge doping data ($\Delta n$) in **Figure 1f** shows that the electron density within Re-MoS$_2$ films initially increases by ~ 7.4·10$^{12}$ e$^-$/cm$^{-2}$ at low dopant concentrations before decreasing as the concentration of Re atoms exceeds ~ 0.5 atom%. This result is consistent with our VBM measurements (**Figure 1e**) and indicates a lack of n-type doping and increased strain in highly doped films.

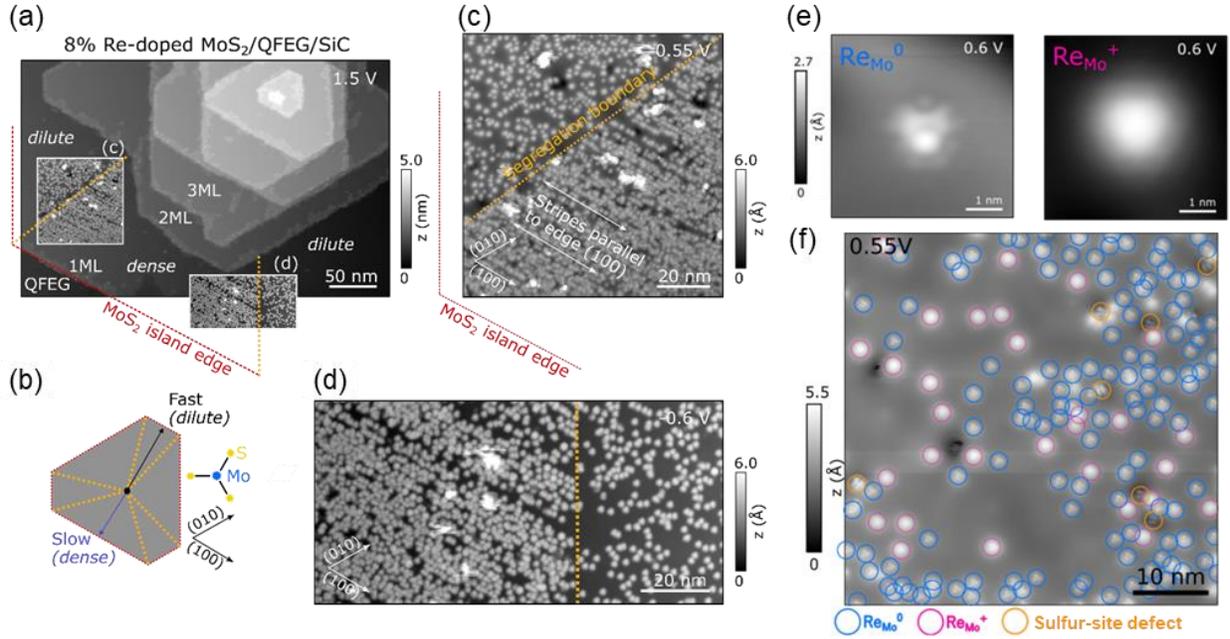

**Figure 2. Rhenium clustering and stripe formation in 8% Re-doped MoS$_2$ on QFEG:** Constant current STM overview image of multilayer Re-doped MoS$_2$ islands. Red and orange dotted lines indicate island edge and segregation boundaries, respectively. (b) Structural model of the MoS$_2$ island with fast (dilute Re concentration) and slow (dense Re concentration) growth facets indicated. (c,d) Constant current STM topography in different areas of the island's monolayer shown in the insets in (a). The segregation boundary is identified from the abrupt change in Re concentration and distribution. e) STM topography of neutral (ReMo$^0$) and positively charged (ReMo$^+$) Re atoms in monolayer Re-MoS$_2$. (f) STM topography highlighting the distribution of neutral (blue circles) and positively charged (magenta circles) Re atoms as well as sulfur-site defects (orange circles) within a monolayer Re-MoS$_2$ film.

The local concentration of Re atoms influences their charge state in Re-MoS$_2$. On QFEG, STM measurements (**Figure 2**) reveal Re$_{Mo}$ substitutional dopants in the neutral and positive charge state, as identified previously.[36] We find that Re dopants in samples with an average doping level of 8 atom% tend to segregate into domains of high (~10 atom%) and low (~3 atom%) concentrations with abrupt transition regions as shown in **Figure 2c,d**. We suspect that the growth kinetics governs the formation of these domains, where the slower growth front accumulates more Re dopants (see model in **Figure 2b**). Additionally, Re dopant distributions may be affected by the edge terminations of Re-MoS$_2$ domains during the growth process.[48] Analyzing the charge state distribution, we find twice as many ionized dopants in low-density areas (**Figure 2e,f**). In high-density regions, more Re atoms tend to be charge neutral if nearby Re atoms are already ionized, indicating an increase in their ionization energy. Interestingly, Re impurities exhibit a preference for aligning in stripes along the (100) direction, particularly on island edges, as observed in the STM topography shown in **Figure 2c**. This stripe-like phase resembles previous reports on W$_x$Mo$_{1-x}$S$_2$ alloying.[49] In this phase, Re atoms often arrange along stripes in the fifth-nearest neighbor configuration (two Mo rows skipped). This is verified by CO-tip nc-AFM imaging in **Figure S6** which reveals the lattice registry of Re atoms in such stripes, highlighted by dashed circles. A corresponding STM image is shown in **Figure S6**, suggesting that the Re stripes may emerge from a pseudo Jahn-Teller distortion of the single Re$_{Mo}^0$ that distorts the local crystal lattice and propagates along the stripe direction.[36]

Density functional theory (DFT) modeling reveals an increased Re dopant ionization energy at small Re-Re separations (**Figure 3**). Dopants in conventional semiconductors, such as phosphorus donors in silicon, are slightly repulsive, as reflected in a +0.05 eV pairing energy[50] (unless exposed near the surface in a nanowire environment).[51] However, for the case of Re in monolayer MoS$_2$, we estimate a Re-Re pairing energy of at least –0.44 eV, i.e., strongly attractive. The strong stabilization of a Re-Re dopant pair is related to strong local distortions. When separated, each Re dopant contributes to a spin-polarized dopant level near the conduction band edge. Following the effective mass approximation, each occupied dopant state can be described by a 2D hydrogenic wavefunction envelope centered on a Re atom consisting of $d_{z^2}$ orbitals inherited from the MoS$_2$ conduction band edge at the K point. When two Re dopants are one lattice constant apart, and each held at a high-symmetry C$_{3v}$ configuration before relaxation, $d_{z^2}$ orbital contributions persist. Since Re dopant ionization energies are on the order of 0.05-0.1 eV[52] for bulk MoS$_2$ and 0.2 eV for monolayers,[53] we expect the stabilization energy of the Re-Re pair without relaxation to be much smaller. Indeed, DFT calculations yield a Re-Re pairing energy of only 0.03 eV for this configuration. However, relaxing the Re pair (**Figure 3c**) results in a pseudo Jahn-Teller distortion similar to a previous report on isolated Re (**Figures 3d and S7**), where a significant $d_{x^2-y^2}$ and d$_{xy}$ mix into the occupied dopant level. The mixing results in the relaxed (distorted) Re pair becoming a deep-level defect, and the effective mass approximation is no longer applicable. The strong Re pair attraction suggests Re-doping MoS$_2$ may be unique from n-doping of conventional semiconductors, with donor deactivation caused by Re cluster formation manifesting at < 1% dopant concentrations.

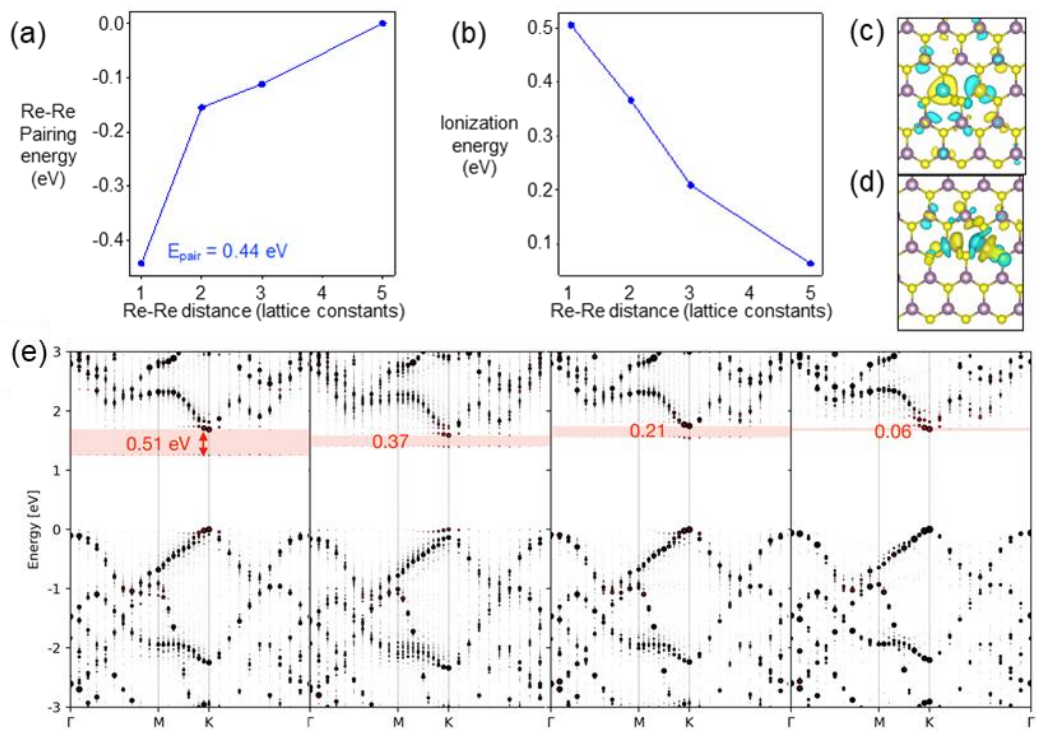

**Figure 3. Impact of clustering on electronic structure of Re in MoS$_2$:** (a) Re dopant pairing is strongly attractive in monolayer MoS$_2$, at 0.44 eV per Re pair. (b) Ionization energy increases for shorter Re-Re separation. Wavefunction near a Re pair (c) before and (d) after allowing the dopant pair to relax into a locally distorted structure. (e) Unfolded band structures of a pair of Re dopant in monolayer MoS$_2$ at separations of 1,2,3, and 5 lattice constants. Trends in Re dopant ionization energies are estimated by monitoring the separation between the donor level and the conduction band edge. The ionization energy (energy difference between defect state in red and CBM) is largest for a nearest neighbor Re-Re pair, at 0.51 eV.

A finite binding energy alone does not guarantee dimer formation since dimerization disfavors configurational entropy. That is, Re-Re dimer formation is only thermodynamically favored when Re-Re attraction (0.44 eV) is stronger than the defect formation energy of an isolated Re atom in the MoS$_2$ lattice.[54] Although evaluating Re dopant formation energies from first-principles requires a Re chemical potential, which is in general unknown, we can estimate the Re dopant formation energy ($E^f$) from the experimentally observed Re concentration $c = 10\%$ using $c = e^{-E^f/kT}$. Assuming a growth temperature of $T = 1200$ K, we estimate a $E^f$ of 0.25 eV, weaker than Re-Re attraction. Thus, we conclude that Re dimer formation is moderately favored in the MoS$_2$ lattice at the growth temperature. We next evaluate whether forming larger Re clusters is thermodynamically favorable. From $E_{mix} = E(Re_xMo_{1-x}S_2) - x\,E(ReS_2) - (1-x)\,E(MoS_2)]$, where we have chosen $x=1/9$ as a close approximation to the experimentally observed doping level of 10%, we obtain a mixing energy per metal atom of 0.11 eV. This moderate energetic preference towards Re clustering (instead of being uniformly distributed) is compensated entropically by $-TS_{mix} = 2k_BT\,[\,x\ln x + (1-x)\ln(1-x)\,] = -0.06$ eV, again using a growth temperature of 1200 K. Thus, a moderate preference of 0.05 eV per metal atom towards Re clustering remains. This result is consistent with the analysis above for Re dimers. However, since this preference is on the same order of thermal energies (~0.10 eV) at 1200 K, we acknowledge that forming large Re clusters versus uniformly distributed dopants also likely depends on local fluctuations of the Re chemical potential.

The calculated ionization energies of Re dopants as a function of Re-Re distance are shown in **Figure 3b.** These ionization energies are estimated by taking the difference (red shaded region in **Figure 3e**) between the pristine band gap and the VBM-defect energy interval, all taken at the DFT level; since we are using only Kohn-Sham levels, we focus on monitoring their trend as the Re pair forms. Donor levels in the unfolded band structures in **Figure 3e** can be identified by the projection onto Re orbitals shown as red dots. The estimated ionization energies increase from 0.06 eV when Re dopants are isolated to 0.37 eV when dopants are two lattice constants apart and then abruptly to 0.51 eV when dopants are one lattice constant apart. Therefore, Re atoms are less likely to ionize when closely neighboring other Re due to the increased ionization energies of clustered dopants. This result supports the lack of n-type doping observed in Re-MoS$_2$ films that exhibit Re clustering at high dopant atom%.

Rhenium incorporation, and its impact on defect formation in MoS$_2$, heavily influences photoluminescence (PL) of MoS$_2$. The PL spectra of undoped and 0.1 atom% Re-MoS$_2$ films collected at room temperature following optical excitation at 445 nm are displayed in **Figures 4a-b**. The film's absorption spectra are shown in **Figure S8**. Non-radiative recombination pathways associated with trion formation quench emission in monolayer MoS$_2$.[55] Therefore, the reduced intensity of the 0.1 atom% film's emission spectrum compared to undoped MoS$_2$ is due to enhanced trion formation in the former. We fit the spectra with two pseudo-Voigt curves centered at ~1.86 and 1.90 eV to determine the contribution of trions (red dashed line) and neutral A-excitons (blue dashed line) to the overall emission.[11, 35] From the intensity ratio of the trion and A-exciton peaks, the electron densities of undoped and 0.1 atom% Re- MoS$_2$ films were found to be

~5.0·10$^{12}$ and 1.2·10$^{13}$ e$^-$/cm$^{-2}$, respectively, using a mass action model (See Supporting Information).[35] The ~7.0·10$^{12}$ e$^-$/cm$^{-2}$ increase in electron density obtained from PL measurements following dilute, 0.1 atom% Re doping is consistent with our Raman and VBM measurements (**Figure 1**). The PL spectrum of a highly doped (3.6 atom%) Re-MoS$_2$ film collected under identical conditions is shown in **Figure 4c**. Unlike the PL spectra of undoped and 0.1 atom% Re-MoS$_2$, the PL spectrum of the 3.6 atom% film exhibits a broad emission peak ~0.2 eV below the neutral A-exciton energy (grey dashed line, **Figure 4c**). The energy separation between this peak and the neutral A-exciton emission is notably more than the separation between trion and neutral A-exciton emission features (~30-40 meV). However, the position of this emission in **Figure 4c** closely matches the calculated energy of Re-related defect states in MoS$_2$ (~1.7 eV) and the indirect bandgap of ReS$_2$. For simplicity, the states that give rise to the

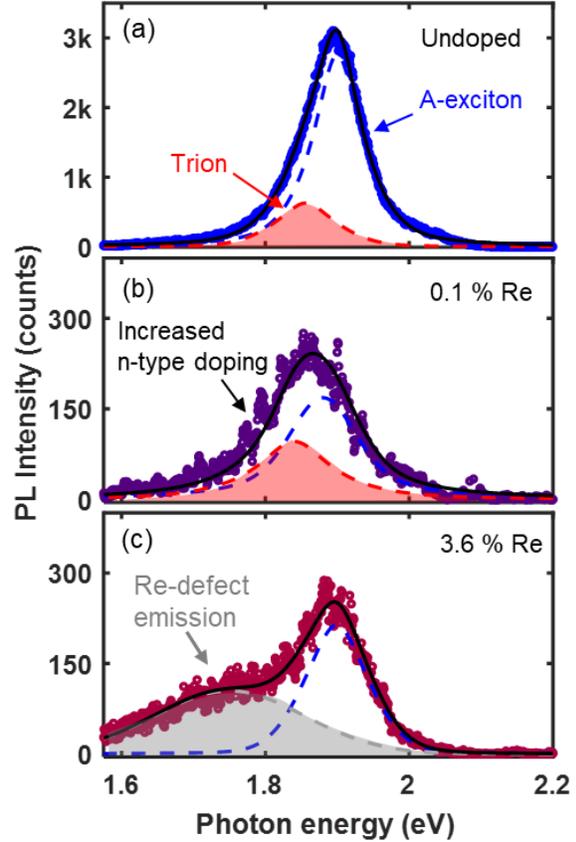

**Figure 4. Emission properties of Re-MoS$_2$ films:** Photoluminescence (PL) spectra of (a) undoped and (b) 0.1 atom%, Re-MoS$_2$ films. The spectra are fit with two pseudo-Voigt curves centered at 1.86 and 1.9 eV to determine the contribution of trions (red dashed line) and A-excitons (blue dashed line) to the PL spectra. (c) PL spectrum of a 3.6 atom% Re-MoS$_2$ film. The spectrum exhibits a broad Re-defect emission band centered at ~1.75 eV (grey dashed line).

~1.75 eV emission band will be referred to as a Re-based defect. The reduced intensity of the 3.6 atom% film's emission spectra compared to undoped MoS$_2$ suggests that the clustering of Re atoms creates defect states that quench emission in highly doped Re-MoS$_2$ films. We note that the overlap between Re-based defect, trion, and A-exciton emission peaks in the PL spectrum of the 3.6 atom% film prevents us from accurately determining carrier densities from the PL spectrum as was done for the undoped and 0.1 atom% Re-MoS$_2$ films. However, the sharpening of the 3.6 atom% sample's emission peak around 1.9 eV compared to the 0.1 atom% film suggests reduced trion formation in the former, in agreement with our VBM and Raman measurements.

Rhenium-based defects trap photogenerated excitons and extend photoluminescence lifetimes in $MoS_2$. PL decay traces of undoped, 0.1, and 3.6 atom% Re-$MoS_2$ films collected near the neutral A-exciton resonance between 1.85-1.9 eV following optical exciton (445 nm, 5 pJ/pulse) are shown in **Figure 5a**. The decay traces presented in **Figure 5a** exhibit an initial, fast component (< 500 ps) and a slow component (> 500 ps) previously assigned to the decay of free excitons and defect-related states [9, 35] We fit the decay traces with bi-exponential functions (see **Table S2** for best-fit parameters) to obtained average exciton recombination lifetimes of 300, 126, and 220 ps for undoped, 0.1, and 3.6 atom% Re-$MoS_2$ films, respectively. Several reports suggest that sulfur-site defects extend PL lifetimes in $MoS_2$.[9,35,56,57] Therefore, we speculate that the extended PL lifetime observed in the undoped $MoS_2$ film arises from sulfur-site defects within the $MoS_2$ lattice. The faster recombination lifetimes of Re-doped samples likely originate from a reduction in these sulfur-site defects (**Figure 1**) and, for the 0.1 atom% Re film, enhanced non-radiative recombination due to increased trion formation.[9]

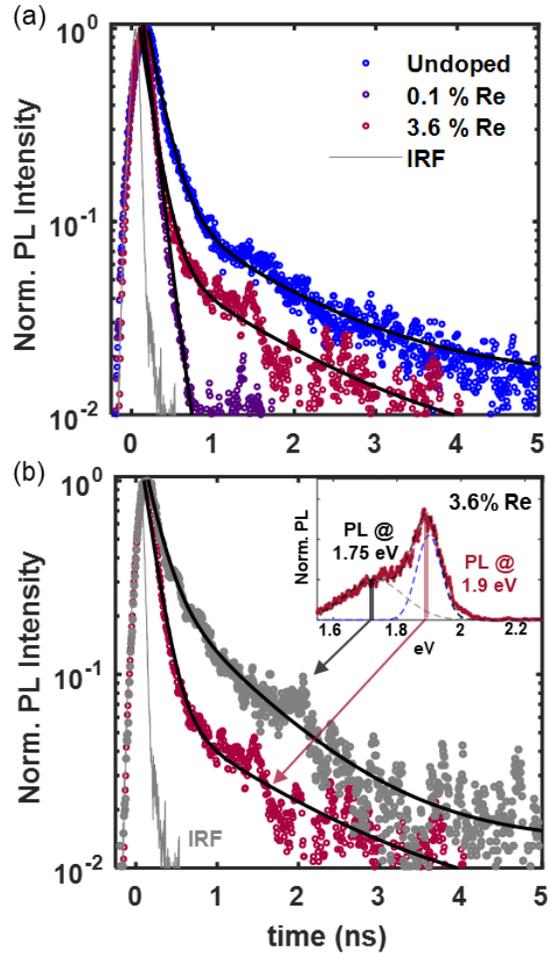

**Figure 5. Influence of Re doping on exciton recombination in $MoS_2$:** (a) Exciton recombination kinetics of undoped, 0.1, and 3.6 atom% Re-$MoS_2$ films measured between 1.85-1.9 eV following optical excitation at 445 nm. (b) Spectrally resolved recombination kinetics of a 3.6 atom% Re-$MoS_2$ film obtained by integrating the film's PL between 1.7-1.75 eV and 1.85-1.9 eV. Inset: Photoluminescence spectra of a 3.6 atom% Re-$MoS_2$ film. The comparison demonstrates that excitons in the 3.6 atom% film relax into Re-defect states that extend exciton lifetimes.

However, unlike the 0.1 atom% sample, the PL decay trace of the 3.6 atom% Re-$MoS_2$ film exhibits a long-lived emission tail at time delays > 500 ps. We conclude that this emission tail does not originate from sulfur-site defects as in the undoped analog because of the marked reduction of sulfur-site defects in the highly doped sample (**Figures 1c & S3**). Instead, long-lived emission in the 3.6 atom% film is due to exciton trapping at Re-based defects. Spectrally resolved PL measurements of the 3.6 atom% Re sample support this conclusion (**Figure 5b**). Namely, the data show that excitons trapped in Re-based defect states (1.7-1.75 eV) have extended lifetimes compared to free excitons probed at 1.9 eV. Thus, the spectral overlap of Re-based defects and free exciton emission gives rise to the long-lived emission tail observed in the decay trace of the 3.6 atom% film collected at 1.85-1.9 eV (**Figures 5a** and **S9**). The above results further support the conclusion that when Re atoms cluster, there is an increase in the ionization energy of the individual Re atoms, thereby leading to defect states that trap photogenerated excitons within the material.

**Conclusion**

We have demonstrated that the local concentration of Re dopants has pronounced effects on lattice strain, charge doping, and exciton trapping in Re-MoS$_2$ monolayers. We show that Re doping reduces sulfur-site defects at doping concentrations from 0.1 to 6 atom%, as observed in Z-STEM. Valance band maximum (VBM) from XPS, Raman, and PL measurements confirm that isolated Re atoms act as n-type dopants in films doped with low concentrations of Re atoms. However, at high dopant concentrations ($\gtrapprox$ 1 atom%), STEM and STM measurements reveal significant Re clustering and stripe-formation throughout the Re-MoS$_2$ lattice. STM measurements and *ab initio* calculations show that the transition from isolated Re atoms to Re clusters at high doping concentrations increase the ionization energy and hence reduce doping from clustered Re atoms. Photoluminescence measurements demonstrate that Re clustering also creates new defect states that trap photogenerated excitons, resulting in broad sub-gap emission. However, while these states may be detrimental to charge doping, impurity-bound excitons from clustered dopants may benefit quantum photonic applications in which such states could act as single photon emitters.[58-60] The results presented here emphasize the need to carefully understand the interplay between local dopant concentrations, carrier doping, and exciton recombination in TMDs when engineering novel devices based on doping 2D semiconductors.

**Experimental**

**MOCVD Growth:** MoS$_2$ films were grown using a home built MOCVD reactor. The growth process for uniform monolayer MoS$_2$ films is detailed in our earlier publication.[35] The concentration of Rhenium dopants in the films was varied by adjusting the flow of H$_2$ carrier gas through a stainless-steel bubbler containing rhenium decacarbonyl [Re$_2$(CO)$_{10}$] powders (99.99% purity, Sigma-Aldrich). The resultant concentration curve, as depicted in **Figure S2**, exhibits a linear relationship between the Re content and the Re$_2$(CO)$_{10}$ flow rate across a wide compositional range (from < 0.1 at.% to > 7 at.%).

**STEM:** Scanning transmission electron microscopy (STEM) images were collected by using a dual spherical aberration-corrected FEI Titan G2 60-300 S/TEM with a high angle annular dark field (HAADF) detector. The parameters for the image collection were a collection angle of 42-244 mrad, camera length of 115 mm, beam current of 40 pA, and beam convergence of 30 mrad.

**X-ray Photoelectron Spectroscopy:**
XPS spectra were collected using a Physical Electronics Versa Probe II tool and a monochromatic Al Kα X-ray source (hν = 1486.7 eV). Samples were measured at high vacuum (<10$^{-6}$ Torr) using a pass energy of 29.35 eV and 0.125 eV energy step. An ion gun and floating electron neutralizer were used to obtain charge neutrality. XPS spectra were charge corrected to C1s spectrum at 284.8 eV.

**Raman Spectroscopy:** Raman spectra were collected using a Horiba Jobin-Yvon LabRam Evolution Raman microscope (Horiba, Edison, NJ). Samples were measured using a 532 nm excitation wavelength.

**STM:** Re-doped (8 atom%) MoS$_2$ samples were grown on QFEG on SiC. Subsequently, the samples were transported through air and annealed at 250-300°C in ultra-high vacuum (~2·10$^{-10}$ mbar). The STM measurements were performed with a commercial low-temperature STM from CreaTec Fischer & Co. GmbH operated at 5 K. STM topographic measurements were taken in constant current mode with the bias voltage applied to the sample. The tungsten tip was prepared on a clean Au(111) surface and confirmed to be metallic.

**Computational Methods**

First-principles density functional theory (DFT) calculations were performed using the generalized gradient approximation with the Perdew-Burke-Ernzerhof (GGA-PBE)[61] exchange-correlation functional using projector augmented wave pseudopotential,[62,63] as implemented in the Vienna Ab initio simulation package (VASP).[64,65] A 5×10 $MoS_2$ supercell with two Re impurities was used in our binding energy calculations. Binding energies obtained from these calculations are lower-bound estimates (i.e. could bind stronger), since they are obtained from a series of total energy calculations with two Re atoms separated by 1–5 lattice constants, where the furthest Re-Re separation structure serves as the reference. We used a Γ-point k-point sampling, a plane-wave expansion energy cutoff of 400 eV, and a force convergence cutoff of 0.01 eV/Å. Band-unfolding method [66] was performed to obtain band structures of the primitive unit cell.

**PL and TRPL characterization**

Steady-state photoluminescence was measured by first focusing the output of a continuous-wave laser onto the sample (Oxxius, LBX-445, 445 nm) using a 40 x 0.75 NA objective. Photoluminescence from the sample was collected with the same objective before being coupled into an optical fiber. A 550 nm dichroic mirror and 600 nm long-pass filter placed before the optical fiber separated the photoluminescence from stray laser light. The fibers output was focused onto the slits of a spectrograph (Princeton Instruments, HRS-300SS, grating 300 grooves/mm) and detected using a back-illuminated CCD camera (Princeton Instruments, PIXIS 400 BR).

Time-resolved photoluminescence (TRPL) was collected by first exciting the sample with 445 nm pulses produced by an optical parametric amplifier (Orpheus-F, Light Conversion) pumped by a 1 MHz Yb:KGW laser (Carbide, Light Conversion Ltd, Vilnius, Lithuania). The same 40 x 0.75 NA objective focused the laser onto the sample and collected the sample's photoluminescence after excitation. An optical fiber focused the photoluminescence onto the slits of a spectrograph before it was detected using a Hamamatsu streak camera (Hamamatsu, C14831-130).

**Acknowledgements**

K.T.M., J.A.R., and J.B.A. acknowledge funding from the U.S. National Science Foundation Major Research Instrumentation program for development of the steady-state and time-resolved PL microscope through award number DMR-1826790. R.T., Y.-C.L, and J.A.R. acknowledge funding from NEWLIMITS, a center in nCORE as part of the Semiconductor Research Corporation (SRC) program sponsored by NIST through award number 70NANB17H041. R.T. and J.A.R. also acknowledge funding from NSF ECCS- 2202280 and NSF DMR-2039351. L.H. and B.S. appreciate funding from the European Research Council (ERC) under the European Union's Horizon 2020 research and innovation program (Grant agreement No. 948243). Y.W. acknowledges startup funds from the University of North Texas and computation allocations from the Texas Advanced Computing Center, the Penn State 2DCC-MIP user program, and NERSC via the Center for Nanophase Materials Sciences user program. For the purpose of Open Access, the author has applied a CC BY public copyright license to any Author Accepted Manuscript version arising from this submission.

**COI Statement**

J. B. A. owns equity in Magnitude Instruments, which has an interest in this project. His ownership in this company has been reviewed by the Pennsylvania State University's Individual Conflict of Interest Committee and is currently being managed by the University.

# Supporting Information

# Influence of Rhenium Concentration on Charge Doping and Defect Formation in MoS$_2$

Kyle T. Munson,[1,†] Riccardo Torsi,[1] Fatimah Habis,[2] Lysander Huberich,[3] Yu-Chuan Lin,[1] Yue Yuan,[4] Ke Wang,[5] Bruno Schuler,[3] Yuanxi Wang,[2] John B. Asbury,[1,4,*] Joshua A. Robinson[1,4,5,6,#]

1. Department of Materials Science and Engineering, The Pennsylvania State University, University Park, Pennsylvania 16802, United States
2. Department of Physics, University of North Texas, Denton, Texas 76203, United States
3. nanotech@surfaces Laboratory, Empa-Swiss Federal Laboratories for Materials Science and Technology, Dübendorf 8600, Switzerland
4. Department of Chemistry, The Pennsylvania State University, University Park, Pennsylvania 16802, United States
5. Materials Research Institute, The Pennsylvania State University, University Park, PA, 16802, USA
6. Department of Physics, The Pennsylvania State University, University Park, Pennsylvania 16802, United States

[†] kmm7044@psu.edu

[*] jasbury@psu.edu

[#] jar403@psu.edu

## Table of Contents

## Section I: Characterization of Re-MoS₂ Monolayers

AFM topography is collected using a Bruker Dimension Icon instrument equipped with a ScanAsyst-Air (k = 0.4 N/m) tip in tapping mode. The topographical image (**Figure S1a**) demonstrates that the MOCVD process produces coalesced, uniform monolayer films on c-sapphire.[1,2] The underlying morphology observed in the AFM image arises from the sapphire substrate's step edges.[3] We evaluated the layer uniformity of a MoS₂ film grown on c-sapphire over a larger area (2500 µm²) using Raman spectroscopy (**Figure S1b**). From the Raman maps, we obtain an average $E^`$ - $A_1^`$ peak distance ($\Delta\omega$) of ~19.5 ± 0.2 cm⁻¹, which closely matches reported $\Delta\omega$ values for monolayer MoS₂ in the literature.[4] Additionally, the spectra exhibit no low-frequency modes associated with multilayer MoS₂.[5]

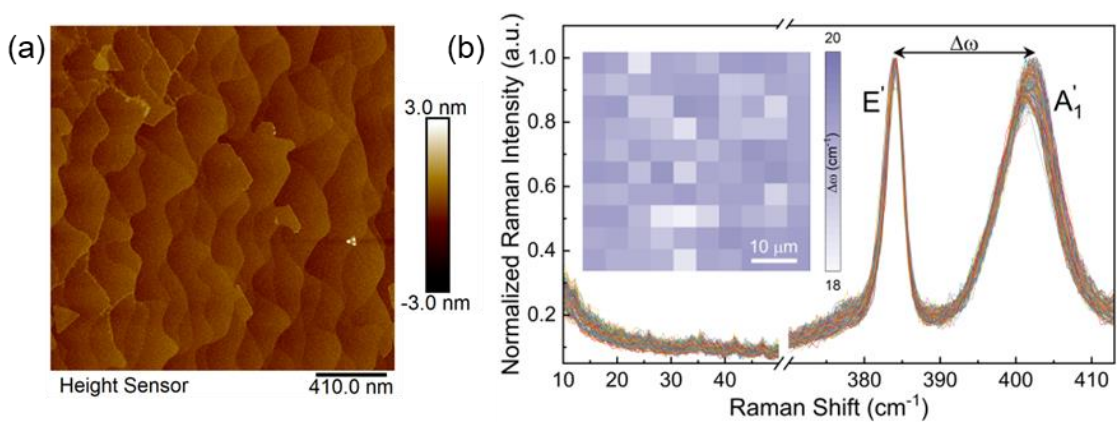

**Figure S1.** (a) AFM topography of an as-grown MoS₂ film on c-sapphire. (b) Raman spectra collected at 100 points across a 2500 µm² region. Inset: Map of $\Delta\omega$ values for a MoS₂ film's $E^`$ - $A_1^`$ modes.

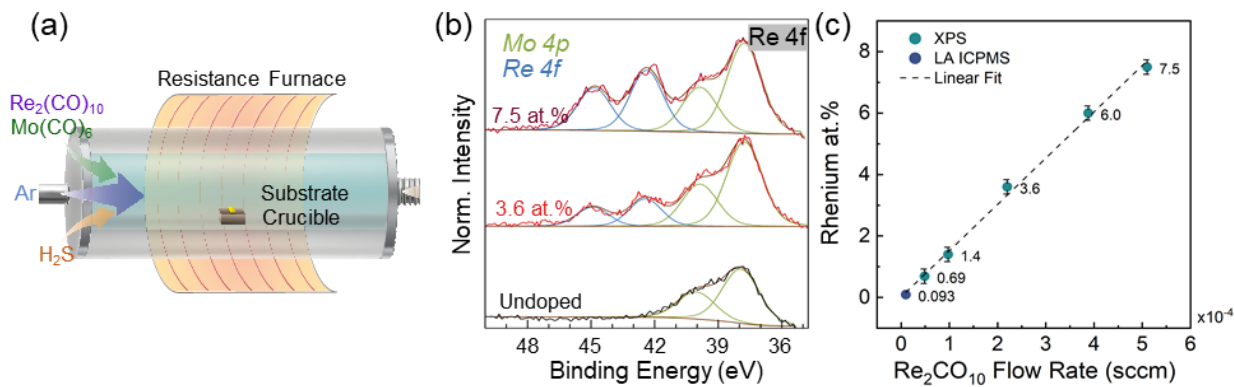

**Figure S2.** (a) Depiction of the MOCVD process for monolayer Re-MoS₂. (b) XPS spectra of undoped, 3.6, and 7.5 atom% Re-MoS₂ films. (c) Re atom% as a function of Re₂(CO)₁₀ flow rate.

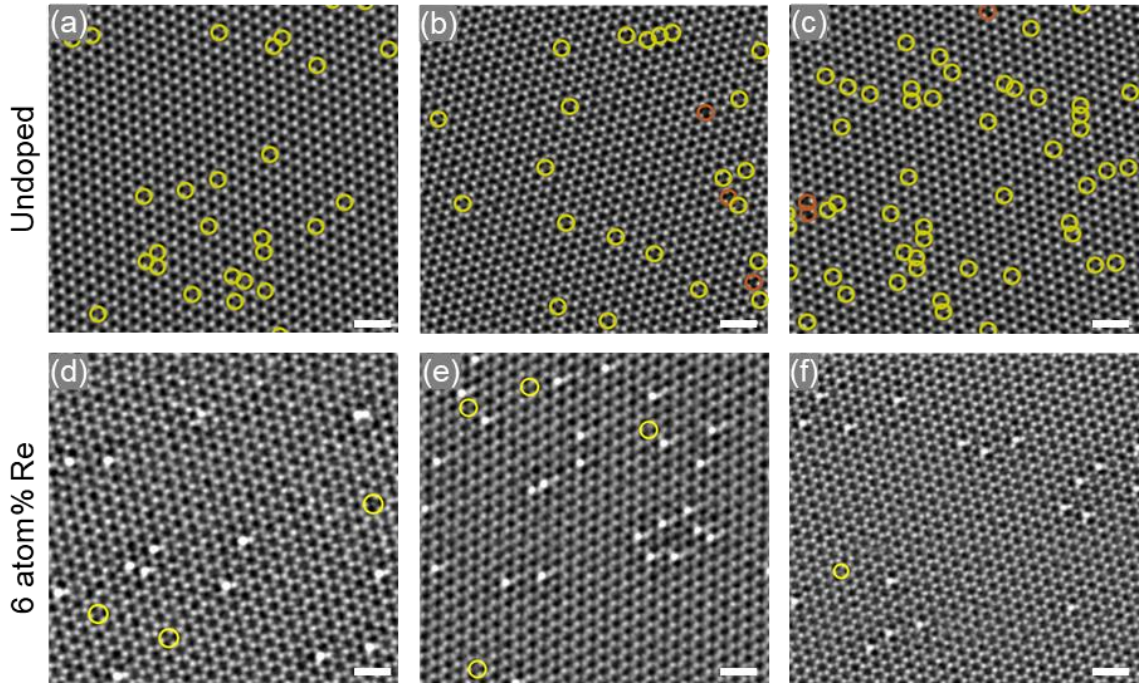

**Figure S3.** Defect density analysis of Z-STEM images for (a-c) undoped and (d-f) 6 atom% Re-MoS$_2$ films collected over a 100 nm$^2$ area. Weak Z-contrast intensity at sulfur sites corresponding to sulfur site-defects (vacancies or light substitutes) and double sulfur-site defects are marked in yellow and orange, respectively. The scale bars are 1 nm.

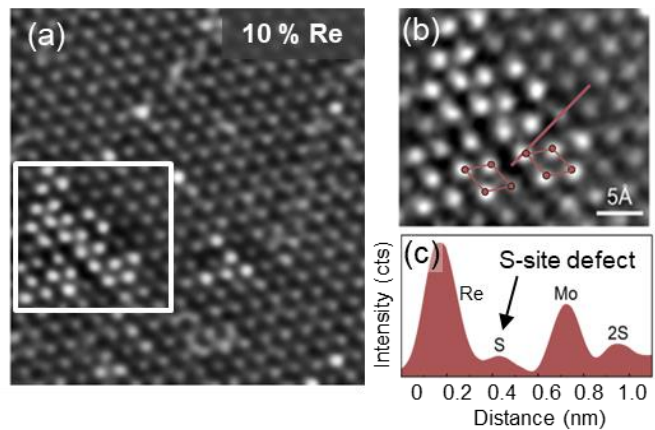

**Figure S4.** (a) Atomic resolution STEM image of a 10 atom% Re-MoS$_2$ film on c-sapphire. (b) STEM image of the same Re-MoS$_2$ film highlighting the presence of Re aggregation. The Re sublattice shown in maroon highlights the crystal structure of ReS$_2$ domains. (c) Line profile depicting the relative Z-intensity of Re, Mo, 2S, and S (sulfur-site defect). Sulfur-site defects are present at the grain boundary between larger Re domains and the MoS$_2$ lattice. This result suggests that the favorable dopant-defect interactions which suppress sulfur-site defects in Re-MoS$_2$ films are reduced at 10 atom% doping concentrations. We speculate that this reduction may arise from increased strain due to phase segregated ReS$_2$ domains.

S3

**Section II: Raman characterization of charge doping and strain in Re-MoS₂ Monolayers**

Eqns. 1-2 in the main text can be expressed in matrix form by,

$$\begin{pmatrix}\Delta\omega_E \\ \Delta\omega_A\end{pmatrix} = \begin{pmatrix}-2\gamma_E\omega_0^E & k_n^E n \\ -2\gamma_A\omega_0^A & k_n^A n\end{pmatrix}\begin{pmatrix}\varepsilon \\ n\end{pmatrix} \quad \text{(Eqn. S1)}$$

Using Eqn. S1, the effects of Re concentration on strain and charge doping density can be determined by constructing a ($\varepsilon$-$n$) map that describes the relationship between these parameters and Raman peak positions (**Figure S5**). **Table S1** lists the Raman peak positions of undoped and Re-MoS₂ films. We note that the $E^`$ and $A_1^`$ frequencies at zero strain and doping are challenging to obtain experimentally. Therefore, we instead compared the strain and carrier concentrations examined in this work to the Raman peak positions of an undoped film. As a result, $\omega_0^E$ and $\omega_0^A$ equal the peak positions of the $E^`$ and $A_1^`$ modes for undoped MoS₂ ($\omega_0^E = E^`_{pristine} = 385.6\ cm^{-1}$ and $\omega_0^A = A^`_{pristine} = 405.5\ cm^{-1}$).

**Table S1.** Raman peak positions for Re-MoS₂ films collected in ten spots across each sample.

| Sample | Mode | $\tilde{\nu}$ (cm⁻¹) | Mode | $\tilde{\nu}$ (cm⁻¹) |
|---|---|---|---|---|
| Undoped | $A_1^`$ | 405.5 ± .1 | $E^`$ | 385.6 ± .1 |
| 0.05 atom% | $A_1^`$ | 403.8 ± .1 | $E^`$ | 385.1 ± .1 |
| 0.1 atom% Re | $A_1^`$ | 404.1 ± .1 | $E^`$ | 385.1 ± .1 |
| 0.5 atom% Re | $A_1^`$ | 404.7 ± .13 | $E^`$ | 384.7 ± .07 |
| 1.4 atom% Re | $A_1^`$ | 404.62 ± .16 | $E^`$ | 384.8 ± .13 |
| 6 atom% Re | $A_1^`$ | 405.2 ± .15 | $E^`$ | 383.8 ± .15 |
| 10 atom% Re | $A_1^`$ | 405.0 ± .15 | $E^`$ | 383.1 ± .25 |

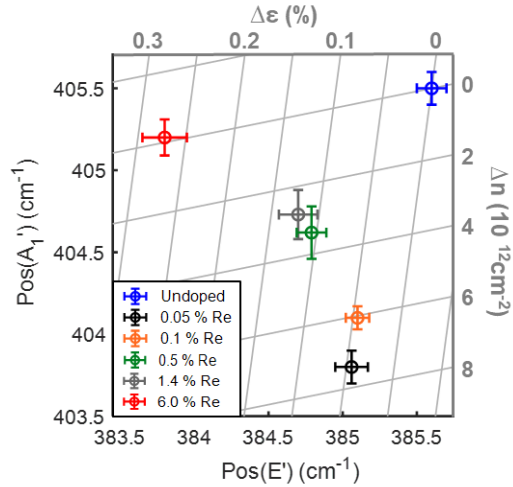

**Figure S5.** Raman-derived strain-charge doping (ε-n) map constructed from the linear relationship between strain, charge doping, and $MoS_2$ Raman peak positions. The error bars indicate the standard deviation of ten Raman measurements collected across the samples.

## Section III: Scanning tunneling microscopy (STM) measurements of Re-$MoS_2$

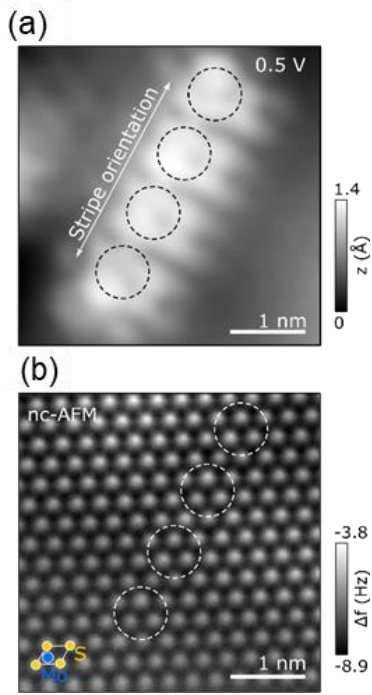

**Figure S6.** (a) STM topography and (b) corresponding nc-AFM image of a four-membered Re strip. The dashed rings indicate the Re position as a guide.

## Section IV: Density functional theory (DFT) calculated Re-MoS$_2$ structures

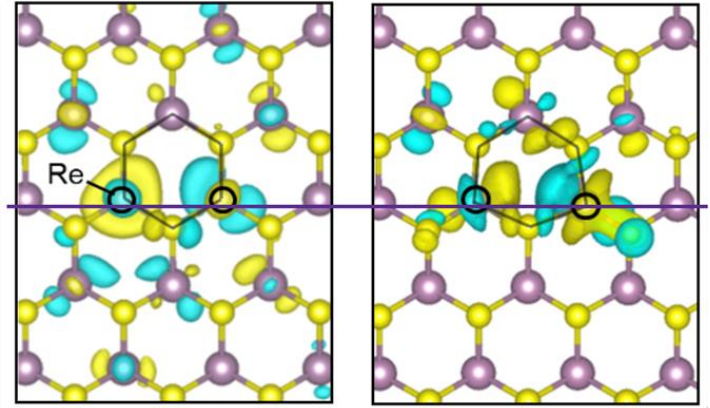

**Figure S7.** Re pair wavefunctions before and after structural relaxation. The structure of the MoS$_2$ lattice around two Re dopants (black circles) is highlighted (black lines). When separated, each occupied dopant state is described by a 2D hydrogenic wavefunction envelope consisting of $d_{z^2}$ orbitals (see main text). After relaxation, the Re pair configuration distorts (right) and $d_{(x^2-y^2)}$ and $d_{xy}$ orbitals mix into the occupied dopant level.

## Section V: Absorption spectra of Re-MoS$_2$ films

We used an Agilent/Cary 7000 spectrophotometer to measure the absorption spectra of undoped, 0.1, and 3.6 atom% Re-MoS$_2$ films directly grown onto transparent c-sapphire. The films have similar optical densities at the pump wavelength (445 nm) used for photoluminescence measurements.

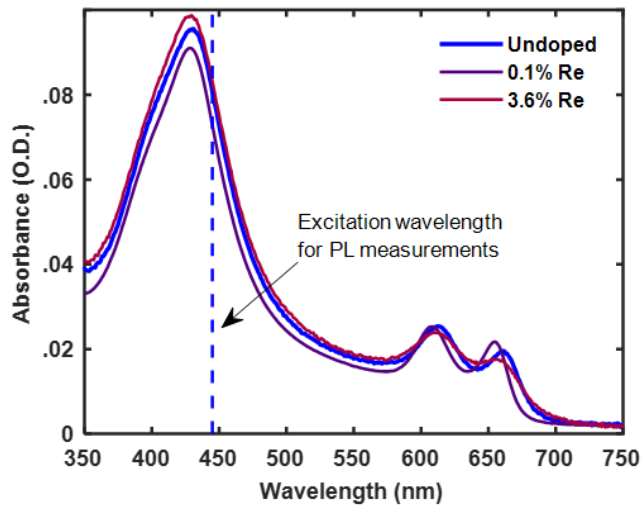

**Figure S8.** Absorption spectra of undoped, 0.1, and 3.6 atom% Re-MoS$_2$ films.

**Section VI: Mass action model to determine electron density from PL spectra**

The electron density within MoS$_2$ films can be estimated from the film's photoluminescence spectra using a mass action model that describes the relationship between neutral excitons, trions, and excess electrons according to

$$\frac{N_{ex}n_{el}}{N_{tr}} = \frac{4m_{ex}m_{el}}{\pi\hbar^2 m_{tr}} k_b T \cdot \exp\left(-\frac{E_b}{k_b T}\right) \qquad \text{(Eqn. S4)}$$

where $N_{ex}$ is the neutral exciton population, $N_{tr}$ is the trion population, n$_{el}$ is the doped electron density, T is temperature, $k_b$ is the Boltzmann constant, $E_b$ is the trion binding energy (~40 meV), and $m_{ex}$, $m_{tr}$, and $m_{el}$ are the effective masses of excitons (0.8 m$_0$), trions (1.15 m$_0$), and electrons (0.35 m$_0$), respectively.[7,8]

The PL intensity of excitons ($I_{ex}$) and trions ($I_{tr}$) is given by

$$I_{ex} \approx \frac{AG\gamma_{ex}}{k_{tr}} \qquad \text{(Eqn. S5)}$$

$$I_{tr} \approx \frac{AG\gamma_{tr}}{\Gamma_{tr}} \qquad \text{(Eqn. S6)}$$

where A is the PL collection efficiency, G is the optical generation rate for excitons, $\gamma_{ex}$ and $\gamma_{tr}$ are the radiative decay rate constants for excitons and trions, $\Gamma_{tr}$ is the total decay rate constant for trions ($\Gamma_{tr} = 0.02\ ps^{-1}$),[7] and $k_{tr}$ is the trion formation rate constant ($k_{tr} = 0.5\ ps^{-1}$).[7]

Using Eqn. S5-6, the electron density of MoS$_2$ films can be estimated from the spectral weight of trion PL according to

$$\frac{I_{tr}}{I_{total}} = \frac{\frac{\gamma_{tr} N_{tr}}{\gamma_{ex} N_{ex}}}{1+\frac{\gamma_{tr} N_{tr}}{\gamma_{ex} N_{ex}}} \approx \frac{4.4 \cdot 10^{-14} n_{el}}{1+4.4 \cdot 10^{-14} n_{el}} \qquad \text{(Eqn. S7)}$$

where $I_{X^-}/I_{total}$ is the trion PL spectral weight. We determined $\gamma_{tr}/\gamma_{ex}$ to be ~0.1 from the intensity ratio of the trion and A-exciton PL curves.

**Section VII: Time-resolved photoluminescence measurements of Re-MoS$_2$ films**

We estimated the PL lifetimes of pristine, 0.1, and 3.6 atom% Re-MoS$_2$ films by fitting normalized PL decay traces with a biexponential function given by

$$D(\tau) = a * exp^{(-t/\tau 1)} + (1-a) * exp^{(-t/\tau 2)} \qquad \text{(Eqn. S8)}$$

where ($\tau 1$) and ($\tau 2$) are short and long lifetime components of the PL decay, and (a) determines the contribution each makes to the overall decay trace. **Table S2** shows the short and long lifetimes, amplitudes, and weighted average lifetimes for the decay traces in Figure 4 of the main text. **Figure S9** displays spectrally resolved PL spectra of a 3.6 atom% Re-MoS$_2$ film collected at several time delays following photoexcitation at 445 nm. The comparison demonstrates that the spectral overlap between Re-defect and free exciton emission gives rise to the long-lived emission tail observed in the decay trace of the 3.6 atom% film collected at 1.9 eV (Figures 4a, main text).

**Table S2.** Biexponential fit parameters for the PL decay traces presented in Figure 4a of the main text.

| Sample | a | $\tau 1$ (ps) | $\tau 2$ (ps) | $\tau_{ave}$ (ps) |
|---|---|---|---|---|
| Undoped | 0.9 ± .01 | 190 ± 6 | 1300 ± 50 | 300 ± 20 |
| 0.1 atom% Re | 0.84 ± .01 | 120 ± 5 | 160 ± 6 | 126 ± 7 |
| 3.6 atom% Re | 0.94 ± .01 | 140 ± 4 | 1450 ± 60 | 220 ± 6 |

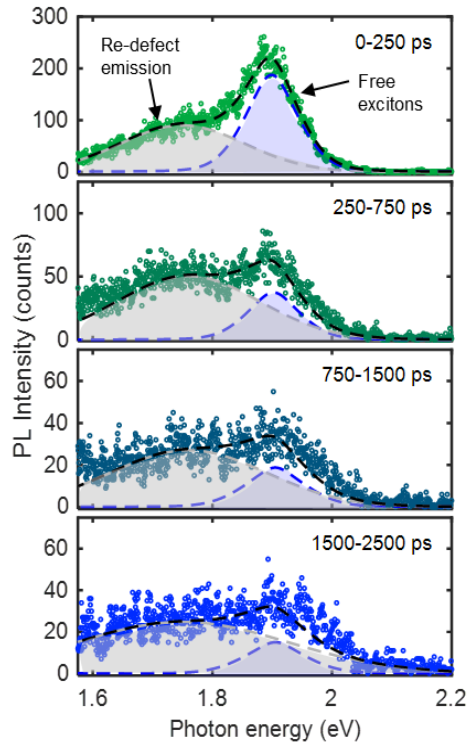

**Figure S9.** Time-resolved PL of a 3.6 atom% Re-MoS$_2$ film obtained by averaging several spectra from early (0−250 ps) to late (150−2500 ps) time delays following 445 nm excitation. The comparison demonstrates that emission from Re-defect states dominates at late time delays and overlaps significantly with emission from free excitons/trions.

# Section VIII: References